\PassOptionsToPackage{table}{xcolor}
\documentclass[sigconf, 10pt]{acmart}

\usepackage{adjustbox}
\usepackage{xspace}
\usepackage{algorithm}
\usepackage{algorithmic}
\usepackage{tikz}
\usepackage{varwidth}

\definecolor{sakura}{cmyk}{0,0.15,0.05,0.05}

\copyrightyear{2026}
\acmYear{2026}
\setcopyright{cc}
\setcctype{by}
\acmConference[ICCAD '26]{IEEE/ACM International Conference on Computer-Aided Design}{November 08--12, 2026}{San Jose, CA, USA}
\acmBooktitle{IEEE/ACM International Conference on Computer-Aided Design (ICCAD '26), November 08--12, 2026, San Jose, CA, USA}
\acmDOI{10.1145/3831252.3833936}
\acmISBN{979-8-4007-2873-0/2026/11}

\newcommand{\name}{{Celty}\xspace}

\newcommand*\circled[1]{\tikz[baseline=(char.base)]{
            \node[shape=circle,fill,inner sep=0.8pt] (char) {\textcolor{white}{#1}};}}

\begin{document}

\title{Celty: SpMSpV GPU Kernel and SIMT Co-Design for Efficient Dual-Sparse LLM Inference}

\author{Ruokai Yin}
\affiliation{%
  \institution{Yale University}
   \city{New Haven}
   \country{USA}
}
\author{Priyadarshini Panda}
\affiliation{%
  \institution{University of Southern California}
 \city{Los Angeles}
   \country{USA}
}

\begin{abstract}
Large Language Models (LLMs) increasingly rely on sparsity to cut inference cost, but most prior work exploits a single sparsity source and targets batched multi-user inference. Dual-sparsity, which pairs unstructured weight pruning with runtime activation sparsity, offers a compelling size–accuracy–latency tradeoff for single-user decoding, but forms a Sparse Matrix–Sparse Vector (SpMSpV) workload that existing GPU kernels handle poorly. We propose \name, a co-designed sparse format, GPU kernel, and SIMT microarchitecture for SpMSpV in LLM inference. \name's Run-Length Compressed CSC (RLC-CSC) format enables vectorized loading of compressed weight columns and exploits both sparsity sources to skip memory accesses, accumulating partial products in shared memory. The \name Sparse SIMT Core then adds a pipelined RLC decoder that removes software index reconstruction and repurposes local register files for conflict-free accumulation, operating on the same compressed representation. The kernel alone achieves up to 2.8$\times$ over cuBLAS; with the Sparse SIMT Core, speedup reaches 5.3$\times$ over cuBLAS at 70\% dual-sparsity.
\end{abstract}

\thanks{This work was supported in part by CoCoSys, a JUMP2.0 center sponsored by DARPA and SRC, the NSF (CAREER Award, Grant \#2312366, Grant \#2318152), the DARPA Young Faculty Award, the DoE MMICC center SEA-CROGS (Award \#DE-SC0023198) and the Global Industrial Technology Cooperation Center (GITCC) program. Correspondence: ruokai.yin@yale.edu}


\maketitle

\section{Introduction}

Large Language Models (LLMs) have demonstrated remarkable capabilities across a wide range of tasks~\cite{touvron2023llama, agarwal2025gpt}, and the vision of personal intelligence—where a capable LLM serves as a dedicated assistant on a user's own local device—is rapidly gaining traction~\cite{li2024personal}. However, deploying increasingly capable LLMs on personal devices faces two fundamental bottlenecks: growing model weights strain limited device memory, and autoregressive decoding is bounded by memory bandwidth when serving a single user. Together, they cap the scale and quality of models that can be practically deployed outside data centers.

Sparsity---removing a subset of model weights---addresses both, shrinking the memory footprint and the data movement required per token. Structured pruning~\cite{men2024shortgpt, zhong2024blockpruner} removes entire blocks or sub-modules and is straightforward to accelerate on GPUs, but degrades accuracy sharply, pushing the community toward unstructured weight pruning~\cite{frantar2023sparsegpt, sun2023simple}.

While unstructured weight sparsity delivers on accuracy and compression, translating it into wall-clock GPU speedup remains a persistent challenge. Its irregular memory access pattern demands specialized kernels~\cite{joo2025coruscant, fan2025spinfer, xia2023flash} and microarchitectural support~\cite{huang2023rm, wang2021dual, joo2025coruscant}, and most such efforts target batched multi-user inference, formulating the problem as Sparse Matrix--Dense Matrix (SpMM) multiplication on tensor cores, an assumption poorly suited to the single-user decoding central to personal-device deployment. And the gains remain modest: Flash-LLM~\cite{xia2023flash} achieves only 1.3$\times$ speedup over cuBLAS at 70\% sparsity on a 4096$\times$4096 layer.

Activation sparsity, identified at runtime through magnitude thresholding~\cite{yin2025duogpt, liu2024training, zhangr}, is a complementary source of sparsity that skips entire weight columns during decoding with minimal accuracy loss, directly addressing unstructured weight sparsity's acceleration weakness for single-user decoding. Because it is input-dependent, however, the full weight matrix must remain in GPU memory, offering no model size reduction.

\textbf{Dual-sparsity}~\cite{yin2025duogpt} resolves this tension by pairing activation sparsity with moderate unstructured weight pruning: the weight sparsity compresses the stored model while activation sparsity drives runtime acceleration, and when both are present, even fewer weights need to be fetched per token. This combination represents a highly promising workload for efficient single-user LLM decoding, formulated as Sparse Matrix–Sparse Vector (SpMSpV) multiplication. Yet existing SpMSpV kernels~\cite{li2020adaptive, ji2022tilespmspv, xu2024ham, yang2015fast} are designed for graph traversal workloads at near-99\% sparsity, and no prior work has investigated efficient GPU kernel support for SpMSpV under the moderate dual-sparsity characteristic of LLM inference.

To bridge this gap, we propose the \name SpMSpV GPU kernel, which employs a Run-Length Compressed CSC (RLC-CSC) sparse format whose strict one-to-one mapping between indices and non-zero values enables efficient vectorized loading of compressed weight columns and scatter-accumulation of partial products in shared memory. Although this kernel already exploits both sparsity sources effectively, profiling reveals that up to 40\% of execution time is consumed by reconstructing row indices from the RLC-CSC format and by shared-memory bank conflicts during scatter-accumulation---bottlenecks that cannot be resolved at the kernel level alone. This motivates a co-design approach: the \name Sparse SIMT Core integrates a lightweight pipelined RLC decoder that eliminates reconstruction overhead entirely in hardware and repurposes local register files as conflict-free partial-product buffers, operating directly on the same RLC-CSC format without any data layout change. Our contributions are summarized as follows:

\begin{enumerate}
\item We propose the \name SpMSpV GPU kernel\footnote{Code is available at \href{https://github.com/RuokaiYin/Celty}{\textcolor{ACMDarkBlue}{https://github.com/RuokaiYin/Celty}}.}, the first GPU kernel designed for dual-sparse LLM inference. We introduce the RLC-CSC sparse format that enables efficient vectorized loading of compressed weight columns and leverage both input and weight sparsity to skip unnecessary memory accesses during single-user decoding, with shared memory employed for partial-product accumulation. The \name GPU kernel alone achieves up to 2.8$\times$ speedup over cuBLAS and 2.4$\times$ over Flash-LLM, the strongest sparse baseline.
\item We propose the \name Sparse SIMT Core, a lightweight micro-architectural enhancement that integrates a 3-stage pipelined RLC decoder to eliminate software-level index reconstruction and repurposes local register files for conflict-free partial-product accumulation. With less than 0.1\% area overhead, the Sparse SIMT Core achieves up to 5.3$\times$ speedup over cuBLAS at 70\% dual-sparsity and up to 2.8$\times$ over Coruscant's sparse tensor core on single-user decoding workloads.
\item We evaluate \name end-to-end on LLaMA-2-13B, demonstrating 2.28$\times$ decode speedup over cuBLAS at 50\% dual-sparsity while maintaining a WikiText-2 perplexity of 7.2, compared to 4.9 for the FP16 dense model. To our knowledge, this is the first work to co-design the sparse format GPU kernel and SIMT microarchitecture for SpMSpV in LLM inference.
\end{enumerate}

\section{Background and Motivation}

\textbf{2.1 Sparsity for LLM Inference.} LLM sparsity techniques span three categories, each offering a different tradeoff among accuracy, model size, and speedup.

\textit{Unstructured weight sparsity.} Training-free pruning methods such as SparseGPT~\cite{frantar2023sparsegpt} and Wanda~\cite{sun2023simple} remove individual weights based on importance scores derived from Hessian estimates or activation magnitudes. The irregular zero pattern preserves model accuracy well but demands dedicated sparse kernels~\cite{joo2025coruscant, fan2025spinfer} for speedup, which remains modest at moderate sparsity~\cite{lu2023dasp}.

\textit{Structured weight sparsity.} Methods such as ShortGPT~\cite{men2024shortgpt} and BlockPruner~\cite{zhong2024blockpruner} remove entire transformer blocks or sub-modules. NVIDIA's sparse tensor cores, introduced in the Ampere architecture~\cite{a100whitepaper}, natively accelerate a finer-grained 2:4 semi-structured pattern, but enforcing this fixed structure still incurs non-negligible accuracy degradation~\cite{frantar2023sparsegpt}. In both cases, the coarse or rigid pattern is straightforward to accelerate~\cite{men2024shortgpt, sandri20252ssp} but limits achievable model quality. 

\textit{Input activation sparsity.} Recent work induces runtime sparsity through magnitude thresholding~\cite{yin2025duogpt, liu2024training} or SVD-based techniques~\cite{zhangr}, enabling column-skipping during decoding with minimal accuracy loss. However, the input-dependent pattern requires the full weight matrix to remain resident in GPU memory.

\textit{Dual-sparsity.} DuoGPT~\cite{yin2025duogpt} combines activation sparsity with moderate unstructured weight pruning, achieving a favorable tradeoff among speedup, model size, and accuracy for single-user decoding. This formulation motivates the SpMSpV workload targeted in this work. Throughout the paper, $x\%$ dual-sparsity refers to $x\%$ input activation sparsity plus $x\%$ unstructured weight sparsity.

\begin{figure}[t]
\centering
\includegraphics[width=0.85\linewidth]{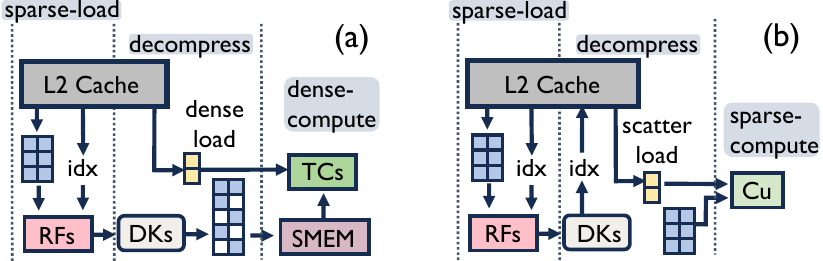}
\vspace{-2.5mm}
\caption{Illustration of (a) prior SpMM kernels. (b) prior SpMV kernels. DK stands for decompression kernel. TC stands for tensor core and Cu denotes the CUDA SIMT cores.}
\vspace{-2.5mm}
  \label{fig:bg_kernels}
\end{figure}

\textbf{2.2 GPU kernels for sparsity in LLMs.} Most GPU kernel research on sparse LLM inference targets the sparse-matrix–dense-matrix (SpMM) operation~\cite{xia2023flash, joo2025coruscant, fan2025spinfer, sgk_sc2020, zheng2022sparta}, which corresponds to batched multi-user decoding on a pruned model. While batching improves GPU throughput, it introduces system-level complexity such as KV-cache management~\cite{kwon2023efficient} and does not address the single-user decoding scenario.

Single-user decoding requires a Sparse Matrix--Dense Vector (SpMV) operation, equivalent to setting N=1 in SpMM. However, state-of-the-art SpMM kernels perform poorly in this regime. These kernels follow a \emph{load-sparse-compute-dense} paradigm~\cite{fan2025spinfer, joo2025coruscant}: sparse data is loaded into registers, decompressed into dense tiles via decompression kernel in shared memory (SMEM), and computed via tensor cores (Figure~\ref{fig:bg_kernels} (a)). While pipelining can hide the decompression latency at large N, at N=1 the tensor core computation is too short to mask the SMEM traffic overhead. Some existing SpMV kernels~\cite{macko2025macko, niu2021tilespmv} avoid this overhead entirely by decoding sparse indices via decompression kernel to scatter-load inputs from global memory and computing directly on SIMT cores (Figure~\ref{fig:bg_kernels} (b)). Other SpMV work \cite{lu2023dasp} leverages tensor core as well to accommodate the computation overhead at very high sparsity regimes.

For dual-sparse workloads, the input vector is itself sparse, transforming the problem into a sparse-matrix–sparse-vector (SpMSpV) operation. The intersection of two irregular sparsity patterns leads to highly unpredictable memory access and workload distribution that existing kernels—designed for a single sparsity source—handle poorly. Prior SpMSpV kernels~\cite{li2020adaptive, ji2022tilespmspv, xu2024ham, yang2015fast} target optimization on graph algorithms such as BFS, at near-99\% sparsity and do not translate to the moderate sparsity levels (30–70\%), which is the standard sparsity regime for LLM inference.

\textbf{2.3 Sparse SIMT Microarchitecture.}  NVIDIA's Ampere architecture introduced sparse tensor cores that natively compute the 2:4 semi-structured format~\cite{a100whitepaper}. To support unstructured sparsity, prior works~\cite{huang2023rm, wang2021dual, joo2025coruscant} have proposed microarchitectural enhancements to the tensor core pipeline. However, all of these target SpMM operations with large
N—either N$\geq2048$~\cite{huang2023rm, wang2021dual} where the workload is strictly compute-bound and justifies the hardware overhead, or N$\geq 8$~\cite{joo2025coruscant} where sufficient pipeline depth hides the decompression latency. In single-user decoding, N is strictly 1, placing the workload firmly in the memory-bound regime, where tensor cores offer no advantage. Rather than enhancing the tensor core, we propose a microarchitectural modification to the SIMT core's CUDA core pipeline tailored to the SpMSpV workload, detailed in Section~\ref{sec:hw}.


\section{Celty SpMSpV GPU Kernel Design}
\label{sec:kernel}
\subsection{\name SpMSpV Formulation}
\begin{figure}[t]
\centering
\includegraphics[width=0.85\linewidth]{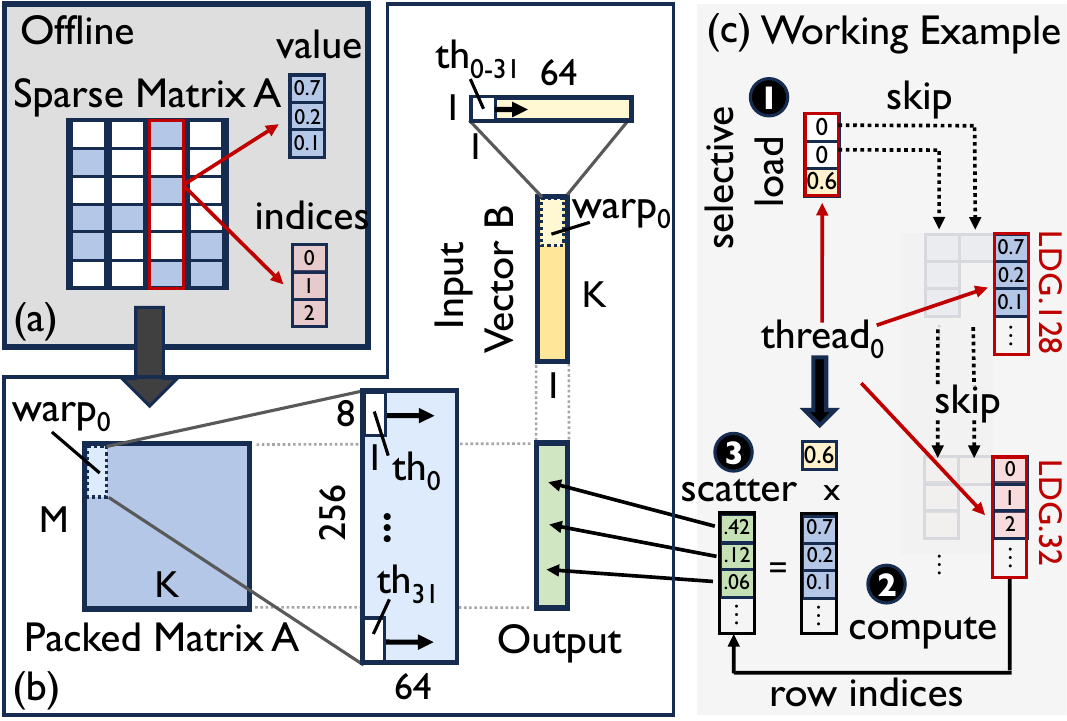}
\vspace{-2.5mm}
\caption{\name SpMSpV kernel design. \textbf{(a)} Offline: weights are packed into RLC-CSC format, which run-length compresses the row indices. \textbf{(b)} Loop ordering: each warp is assigned a tile of 256 rows (M) and iterates over 64 columns (K). \textbf{(c)} Runtime example: zero-valued inputs trigger a column skip; for non-zero inputs, weights and indices are vector-loaded and scatter-accumulated into the output.}
\vspace{-4mm}
  \label{fig:spmsv_dataflow}
\end{figure}

The central design goal of the \name SpMSpV kernel (Figure \ref{fig:spmsv_dataflow}) is to exploit input sparsity to skip loading the corresponding weight columns from global memory, thereby reducing memory traffic and latency. To achieve this, each thread must first inspect the input activation value $b$ before deciding whether to issue weight-fetch instructions. This introduces a critical design consideration: if threads within a warp fetch different $b$ values, the unstructured input sparsity pattern causes different threads to take different control-flow paths, resulting in warp divergence that negates any speedup. To avoid this, all 32 threads within a warp share the same input value $b$, and each warp processes multiple rows of $A$ simultaneously. At each iteration, the warp loads $b$ and checks whether it is zero; if so, the entire warp skips to the next column without issuing any memory access.

Essentially, this resembles a split-K formulation~\cite{hoque2024accelerating} where the K-dimension is distributed across warps. However, increasing K-parallelism also increases the cost of merging partial sums—either through atomic additions to global memory or a separate reduction kernel launch.
To balance parallelism against atomic merge overhead, \name assigns each warp a 2D tile (Figure~\ref{fig:spmsv_dataflow}(b)): the M-dimension is partitioned across warps, each covering 256 rows, while each warp iterates over 64 columns along K in an inner loop. We use 128-bit vectorized loads~\cite{luitjens2025vectorized} for data loading: at FP16 precision, each thread loads 8 consecutive weights (16 bytes), so each 32-thread warp covers 256 rows. To guarantee alignment for vectorized access, we apply reverse offset memory alignment (ROMA) following~\cite{macko2025macko}.

\subsection{\name Sparse Format}
With the kernel formulation established, we now describe the sparse format designed to complement it. Bitmap-based formats are a popular choice for storing unstructured pruned weights~\cite{joo2025coruscant, fan2025spinfer}, offering better compression than traditional Compressed Sparse Row (CSR) or Coordinate List (COO). However, bitmaps are fundamentally incompatible with our vectorized-loading strategy. The root issue is that the number of set bits in a bitmap chunk is unpredictable: when a thread loads 32 bitmap bits, the number of corresponding non-zero values varies, making it impossible to issue a fixed-width vectorized load for the values. Allowing only threads with sufficient non-zeros to issue vectorized loads reintroduces warp divergence.

We instead adopt a Compressed Sparse Column (CSC)-based format, which provides a strict one-to-one mapping between indices and non-zero values—exactly matching the vectorized load granularity. Standard CSC uses a 32-bit row index per non-zero entry, yielding poor compression at the moderate sparsity levels (30–70\%) typical of LLM inference~\cite{joo2025coruscant, fan2025spinfer}. To address this, we compress each row index using
n-bit run-length coding (RLC)~\cite{chen2016eyeriss, chen2019eyeriss, han2016eie, macko2025macko}, where each RLC value encodes the number of zeros preceding the corresponding non-zero entry. Gaps exceeding $2^n-1$ are handled by inserting zero-valued padding entries. For example, the sparse column $[0,0,4,0,0,0,0,2,8]$ is encoded with 2-bit RLC as a value array $[4,0,2,8]$ and a run-length array $[2,3,0,0]$. Figure~\ref{fig:spmsv_dataflow}(a) illustrates the offline packing process, where each column's non-zero values and their RLC-CSC indices are stored contiguously for vectorized access.

\begin{figure}[t]
\centering
\includegraphics[width=0.85\linewidth]{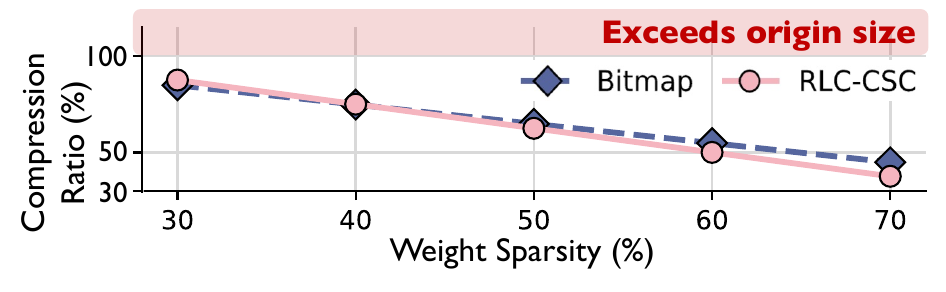}
\vspace{-3mm}
\caption{Compression ratio of bitmap vs. RLC-CSC across different sparsity on a 4096$\times$4096 layer.}
\vspace{-5mm}
\label{fig:compression_efficiency}
\end{figure}

\name uses 4-bit RLC-CSC, achieving approximately 63\% compression efficiency on a 50\% sparse 4096$\times$4096 layer. Although bitmaps offer a theoretically lower bound of  $\sim56\%$ compression efficiency at the same sparsity, their deployment-time compression efficiency is $\sim64\%$ because of the value-index decoupling discussed above. 
Figure~\ref{fig:compression_efficiency} compares the two across sparsity levels: RLC-CSC is comparable to the bitmap implementation of~\cite{joo2025coruscant} at deployment time while additionally being compatible with vectorized loading, which is the property the bitmap cannot provide.

\begin{algorithm}[t]
\caption{Intra-warp prefix sum over RLC run-lengths}
\label{alg:reconstruct}
\begin{adjustbox}{max width=\columnwidth,center}
\begin{varwidth}{2\columnwidth}
\begin{algorithmic}[1]
\STATE \textbf{Inputs:} $lane$ and $\mathbf{rlc\_arrays}$
\STATE $size \leftarrow len(\mathbf{rlc\_arrays})-1$
\STATE Initialize $sum\leftarrow 0$,  $all\leftarrow$ \texttt{0xFFFFFFFF}
\STATE Initialize $\mathbf{prefix}[0:size]\leftarrow 0$
\FOR{$i \leftarrow 0$ \TO $size$}
    \STATE $sum \leftarrow sum + \mathbf{rlc\_arrays}[i]$
\ENDFOR
\FOR{$i \leftarrow 0$ \TO $\log_2 32-1$}
\STATE $pos \leftarrow 2^{i}$
\STATE $prev\_sum\leftarrow$ \textsc{shfl\_up\_sync}$(all, sum, pos)$
    \IF{$lane \ge pos$}
    \STATE $sum \leftarrow sum + prev\_sum$
    \ENDIF
\ENDFOR
\FOR{$i \leftarrow size$ \textbf{downto} $0$}
\STATE $\mathbf{prefix}[i]\leftarrow sum$
\STATE $sum \leftarrow sum - \mathbf{rlc\_arrays[i]}$
\ENDFOR
\STATE \textbf{Outputs: }$\mathbf{prefix}$ \COMMENT{row index adds the entry's ordinal; see code}
\end{algorithmic}
\end{varwidth}
\end{adjustbox}
\end{algorithm}

\subsection{SMEM-Based Scatter-Accumulation}

As discussed, when a warp encounters a non-zero input activation, it must compute and accumulate partial products from the corresponding weight column into the output vector. Figure~\ref{fig:spmsv_dataflow}(c) illustrates this process with a working example for a single thread. The key challenge is determining the correct output row for each loaded weight. Each thread loads 8 weights along with 8 RLC-CSC indices, but because these indices are run-length encoded, threads within a warp must cooperate to reconstruct the absolute row positions via an intra-warp prefix sum over the run-lengths (Algorithm~\ref{alg:reconstruct}); the absolute row index follows by adding each entry's ordinal. The reconstructed indices then serve as scatter addresses for accumulating partial products. 

Naively writing partial sums directly to global memory incurs severe overhead from atomic operations, write contention, and high access latency. To mitigate this, each thread block accumulates into a local shared-memory (SMEM) buffer, which is flushed to global memory only after all assigned columns have been processed. This reduces the frequency of expensive global writes, yielding on average 14.4$\times$ speedup over direct global-memory accumulation on the A5000 GPU. However, SMEM bank conflicts during scatter-accumulation and the inherently higher access latency compared to register files limit the achievable performance, motivating the hardware-level solution presented in Section~\ref{sec:hw}.

\section{Celty Sparse SIMT Microarchitectural Enhancements}
\label{sec:hw}

Profiling the SMEM-based \name SpMSpV kernel reveals two performance bottlenecks that cannot be resolved at the software level: SMEM bank conflicts during scatter-accumulation and the overhead of reconstructing row indices from RLC-CSC (Algorithm \ref{alg:reconstruct}). We address each with a targeted microarchitectural enhancement to the GPU's SIMT core.

\subsection{Register-File-Based Accumulation}
\begin{figure}[t]
\centering
\includegraphics[width=0.75\linewidth]{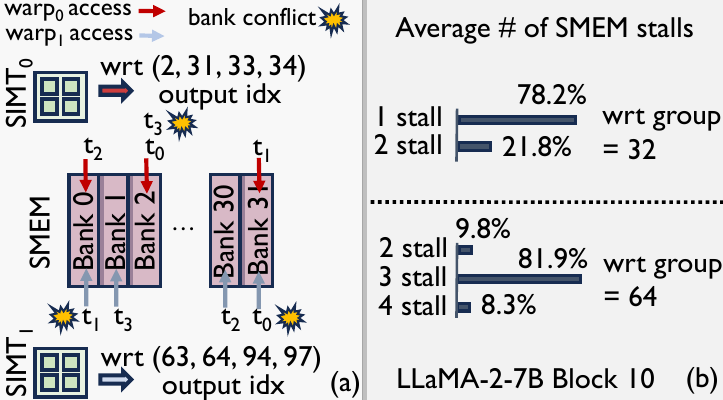}
\vspace{-2.5mm}
\caption{
(a) Shared-memory bank conflicts during scatter-accumulation: threads from different SIMT cores targeting the same bank serialize writes. (b) SMEM stall distribution profiled at 50\% dual-sparsity.}
\vspace{-2.5mm}
\label{fig:bank_conflict}
\end{figure}

In the SMEM-based kernel, scatter-accumulation writes partial products to a shared-memory buffer indexed by the reconstructed row positions. Because a single SMEM bank group (32 banks) is shared across four SIMT cores, concurrent write requests from multiple warps frequently collide on the same bank, serializing the writes and stalling the pipeline. Figure~\ref{fig:bank_conflict}(a) illustrates this mechanism, and Figure~\ref{fig:bank_conflict}(b) quantifies the impact: on a representative LLaMA-2-7B layer at 50\% dual-sparsity, over 90\% of iterations incur 3 or more stall cycles when the write group size reaches 64.

To eliminate this bottleneck, we repurpose the abundant local register files within each SIMT core as partial-product buffers. The output vector is partitioned into four equal segments, one per SIMT core within an SM sub-partition. For example, for $M=5120$, each SIMT core reserves 1280 registers as a local accumulation buffer. During computation, after the RLC-CSC indices are decoded into row positions, each thread writes its partial products directly to the register file at the corresponding output position, bypassing SMEM entirely. These registers are reserved prior to kernel launch and shared among threads within each SM sub-partition\footnote{Details on inter-thread register sharing can be found in prior work on register-file virtualization, such as~\cite{jeon2015gpu}.}.

Since register-file access has both lower latency and higher bandwidth than SMEM, this approach eliminates bank-conflict stalls and reduces accumulation latency. As demonstrated in our ablation study (Section~\ref{sec:ablation}), even partial register allocation (e.g., 50\%) yields significant speedup at low dual-sparsity where SMEM pressure is highest, while full register-resident accumulation provides up to 2.40$\times$ improvement over the pure SMEM baseline. Importantly, the register-accumulation ratio can be tuned to balance performance against occupancy constraints, making the approach practical across different layer dimensions and sparsity levels. In the rare case that a target output position falls outside the locally partitioned register range, SMEM serves as a fallback buffer to ensure correctness; in practice, our profiling shows that such spills occur negligibly across all evaluated workloads.

Here we additionally provide an analysis on register accumulation's impact on the occupancy. From compiled register usage (\texttt{ptxas -v}), on the A5000 (sm\_86, 65536 registers/SM, 48 max warps/SM), the \name kernel uses 40 registers/thread (1280/warp), allowing the full 48 warps/SM. Reserving the registers for accumulation reduces the budget: for M=5120, 5120 reserved registers leave 60416, supporting 47 warps (97.9\%, a 2.1\% drop in occupancy). Across all evaluated layer sizes the reduction stays under 7\%.

\subsection{Hardware RLC Decoder}

\begin{figure}[h]
\centering
\includegraphics[width=0.85\linewidth]{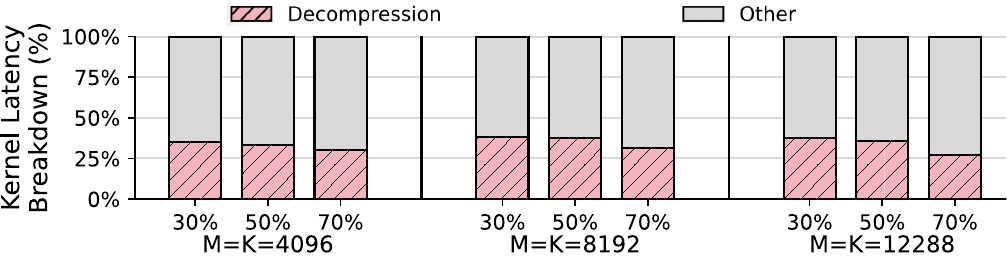}
\vspace{-2.5mm}
\caption{Percentage of RLC-CSC reconstruction latency in total kernel execution time on A5000 across different dual-sparsity levels and layer shapes.}
\vspace{-2.5mm}
\label{fig:decompression_latency}
\end{figure}

While the RLC-CSC format reduces index storage and enables efficient vectorized loading, current GPUs do not natively support it. The kernel must therefore perform intra-warp prefix-sum reconstruction (Algorithm~\ref{alg:reconstruct}) at every iteration before scatter-accumulation can proceed. Notably, this cost is largely masked by SMEM stalls in the baseline kernel, but once register-file accumulation eliminates the SMEM bottleneck, reconstruction emerges as the dominant remaining overhead. As shown in Figure~\ref{fig:decompression_latency}, it can occupy up to 40\% of total kernel execution time, representing a substantial optimization opportunity.

\vspace{-2.5mm}

We propose to eliminate this overhead through a lightweight microarchitectural addition to the SIMT core: a hardware RLC decoder that reconstructs absolute row indices directly from the RLC-CSC stream, removing the need for software prefix-sum instructions and intra-warp shuffle communication. The modification is minimal and allows the SIMT core to operate in both its original dense mode and the \name sparse mode.

\begin{figure}[h]
\centering
\includegraphics[width=0.8\linewidth]{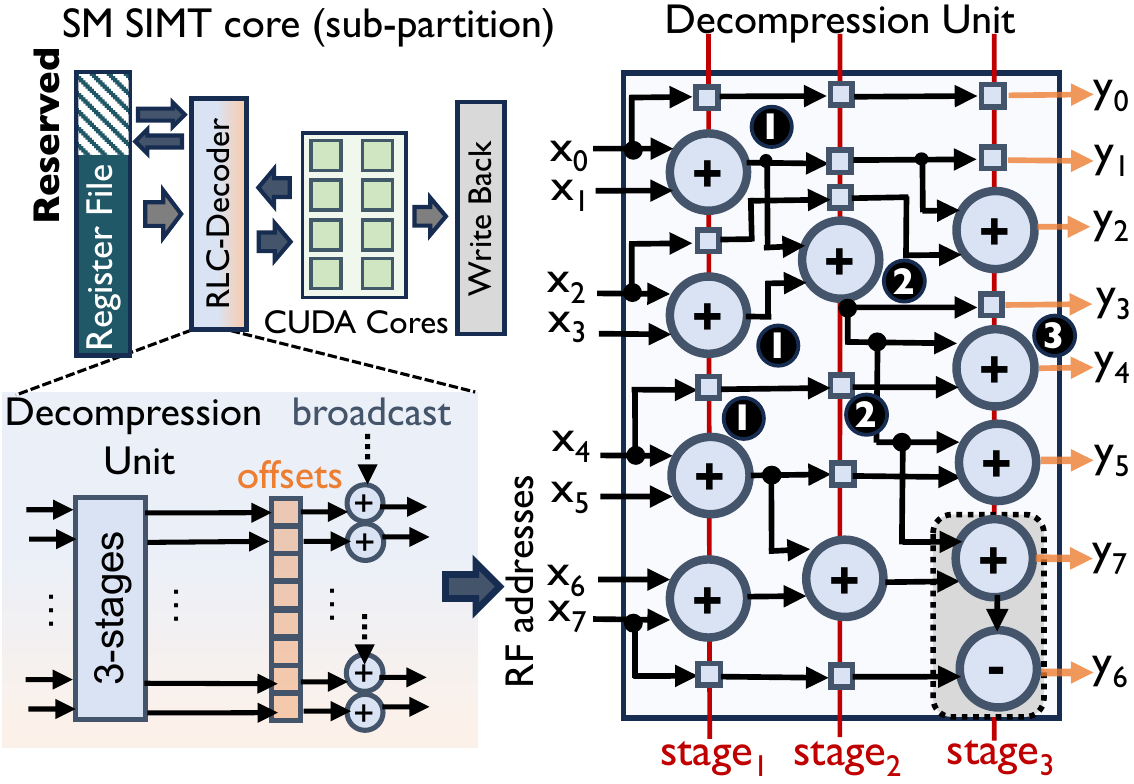}
\vspace{-2.5mm}
\caption{\name Sparse SIMT Core microarchitecture. The RLC decoder reconstructs row indices; the output offsets address the register-file accumulation buffer.}
\vspace{-3.5mm}
  \label{fig:celty_hw}
\end{figure}

\begin{figure*}[t]
\centering
\includegraphics[width=0.9\linewidth]{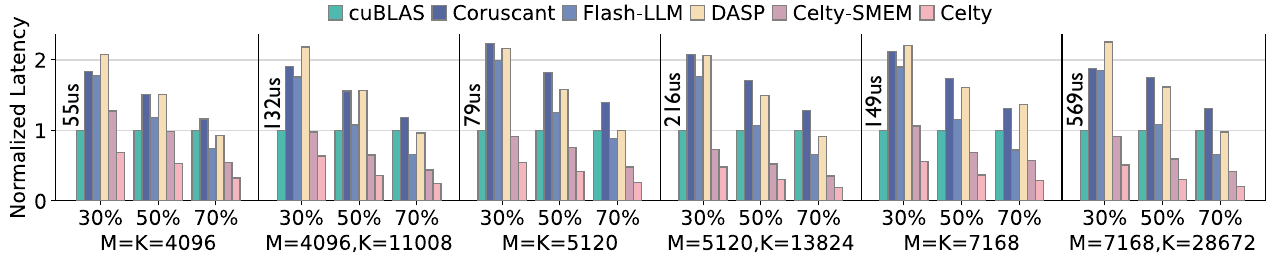}
\vspace{-2.5mm}
\caption{Comparison of kernel latency across workloads and sparsity.}
  \label{fig:results_kernel_latency}
\end{figure*}

Figure~\ref{fig:celty_hw} illustrates the proposed RLC decoder, a 3-stage pipelined unit that ingests 8 RLC-CSC indices per cycle. Once the pipeline is full, 8 reconstructed row indices are produced every cycle. The pipeline implements a modified parallel prefix adder tree that computes cumulative sums of the run-lengths, equivalent to Algorithm~\ref{alg:reconstruct} but executed entirely in hardware. The seventh output ($y_6$) is back-calculated via a subtractor in parallel with the final adder-tree stage, eliminating the need for an extra forward-carry stage.
To illustrate, the markers \circled{1}--\circled{3} trace the computation of $y_4$: stage one (\circled{1}) computes partial sums $x_0+x_1$ and $x_2+x_3$ while buffering $x_4$; stage two (\circled{2}) aggregates these into a block sum $\sum_{i=0}^3 x_i$ with $x_4$ still buffered; stage three (\circled{3}) adds the forwarded $x_4$ to produce the final index $y_4$.

To maximize thread utilization, the decoder ingests its 8 inputs from different thread lanes rather than from a single thread. This requires the offline RLC compression stage to interleave non-zero values and their RLC-CSC indices across threads—for instance, thread 0 loads entries at positions $[0,32,64,96,128,160,192,224]$ in the non-zero sequence. Because scatter-accumulation is order-independent, interleaving does not affect correctness. The reconstructed offsets from the decoder are added to a broadcasted base pointer to form the register-file address for the current partial product, which is then fetched and forwarded to the FP ALU for the fused multiply-accumulate operation. Together with the kernel design in Section~\ref{sec:kernel}, the Sparse SIMT Core completes the \name co-design stack, operating end-to-end on the same RLC-CSC format from storage through computation.

\section{Experimental Setup}

\textbf{Workloads.} We evaluate on linear layer dimensions drawn from three widely used LLM models: 4096$\times$4096 and 4096$\times$11008 (LLaMA-2-7B \cite{touvron2023llama}), 5120$\times$5120 and 5120$\times$13824 (LLaMA-2-13B \cite{touvron2023llama}), and 7168$\times$7168 and 7168$\times$28672 (OPT-30B \cite{zhang2022opt}).

\textbf{Baselines.} We compare against two state-of-the-art SpMM kernels—Coruscant~\cite{joo2025coruscant} and Flash-LLM~\cite{xia2023flash}—and one SpMV kernel, DASP~\cite{lu2023dasp}, alongside dense cuBLAS. All kernels are compiled with CUDA Toolkit 12.6 and GCC 12.2, and evaluated on a single NVIDIA A5000 GPU (Ampere, compute capability 8.6, 24\,GB GDDR6). Since prior sparse kernels exploit only weight sparsity, we apply dual-sparsity (both input and weight) to \name and weight-only sparsity to all baselines, ensuring matched weight sizes across all kernels. For the two SpMM baselines, the
N dimension is padded to 8 to satisfy tensor-core alignment requirements.

\textbf{Hardware simulation and synthesis.} For the \name Sparse SIMT Core, we estimate GPU-level performance following the methodology of VectorSparse~\cite{zhu2019sparse} and Coruscant~\cite{joo2025coruscant}: we implement a modified version of the \name kernel that removes the RLC reconstruction instructions and replaces SMEM-based scatter-accumulation with direct register-file access, while retaining the standard \texttt{fmaf\_rn} computation. This modified kernel is executed on A5000 GPU, and the measured latency serves as an estimate of the performance achievable with the proposed \name Sparse SIMT Core enhancements. We synthesize the RTL implementation of the RLC decoder using Synopsys Design Compiler at 400\,MHz targeting 32\,nm technology and scale to 7\,nm using the methodology of~\cite{stillmaker2017scaling}.

\section{Evaluation}
\subsection{Kernel Performance and Analysis}

Figure~\ref{fig:results_kernel_latency} reports kernel latency across all
evaluated layer dimensions and sparsity levels. The SMEM-based \name kernel (Section~\ref{sec:kernel}) consistently outperforms the tensor-core baselines Coruscant (SpMM), Flash-LLM (SpMM), and DASP (SpMV) across all workloads and sparsity levels, as each requires an additional decompression stage whose SMEM traffic is not amortized at N=1. Against dense cuBLAS, \name is slightly slower at low sparsity on smaller layers (roughly 1.2$\times$ at 30\% for M=K=4096) but delivers speedup from 50\% dual-sparsity onward for most shapes, averaging 1.49$\times$ and 2.19$\times$ at 50\% and 70\% and peaking at 2.80$\times$ on the 5120$\times$13824 layer at 70\%. On that layer it reaches 2.41$\times$ over Flash-LLM, 2.82$\times$ over DASP, and 2.84$\times$ over Coruscant, with margins largest at 30\% where the baselines' decompression overhead is least amortized. The advantage stems from dual-sparsity's lower effective memory traffic, though gains remain bounded by the SMEM bank-conflict and RLC reconstruction overheads discussed above.

With the \name Sparse SIMT Core enhancements, both bottlenecks are resolved in hardware. Averaged across all evaluated layer shapes, \name achieves 1.78$\times$, 2.73$\times$, and 4.06$\times$ speedup over dense cuBLAS at 30\%, 50\%, and 70\% dual-sparsity, respectively. The maximum speedup reaches 5.27$\times$ on the 5120$\times$13824 layer at 70\% dual-sparsity.


\subsection{End-to-End Model Performance}

\begin{figure}[t]
\centering
\includegraphics[width=0.8\linewidth]{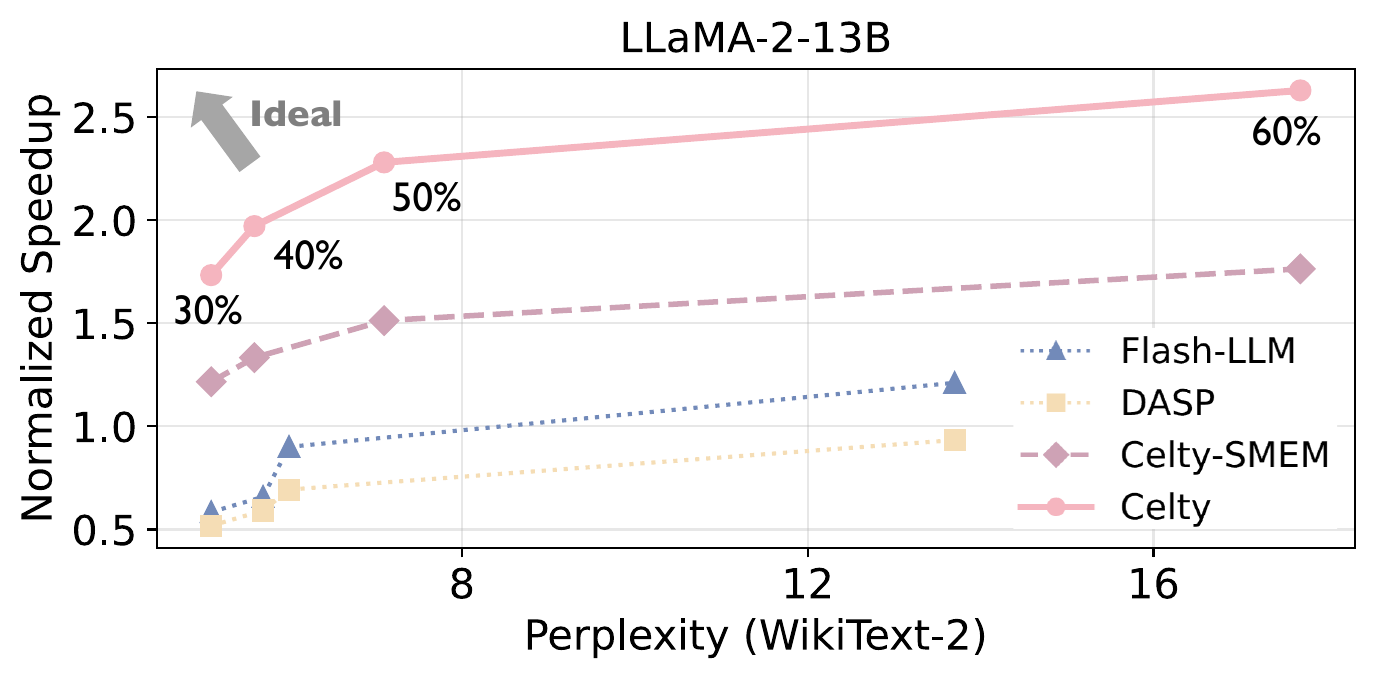}
\vspace{-3.5mm}
\caption{End-to-end decode latency vs.\ perplexity (WikiText-2) under different dual-sparsity configurations. Each point represents different sparsity.}
\vspace{-4.5mm}
\label{fig:end2end}
\end{figure}

To evaluate the practical impact of \name beyond individual layers, we measure end-to-end single-token decode latency on LLaMA-2-13B with 161 prefill tokens~\cite{sharegpt2023} under varying sparsity configurations, reporting the corresponding WikiText-2 \cite{merity2016pointer} perplexity in Figure~\ref{fig:end2end}. For Flash-LLM and DASP, models are pruned using Wanda~\cite{sun2023simple}; for \name, we use DuoGPT's~\cite{yin2025duogpt} dual-sparsity framework. All configurations use FlashAttention~\cite{dao2022flashattention} for the attention layers.

As shown in Figure~\ref{fig:end2end}, \name consistently improves end-to-end decode latency over dense cuBLAS across all evaluated accuracy–sparsity tradeoff points. In contrast, DASP and Flash-LLM fail to achieve speedup below 50\% sparsity despite operating in a similar perplexity range. At 50\% dual-sparsity, \name reaches 2.28$\times$ end-to-end speedup while maintaining a WikiText-2 perplexity of 7.2, compared to 4.9 for the FP16 dense baseline. At higher sparsity, \name reaches up to 2.63$\times$ speedup with increased but bounded perplexity degradation. These results confirm that dual-sparsity, paired with an efficient SpMSpV kernel and hardware support, delivers meaningful model-level latency reduction for single-user decoding.

\subsection{\name Sparse SIMT Core Evaluation}

\textbf{Area overhead.} Table~\ref{tab:area_power} reports the cumulative area overhead of the \name Sparse SIMT Core scaled across GPU generations by the number of SIMT cores per chip. The overhead remains minimal across all generations, reaching at most 0.08\% of additional area on Hopper. For comparison, we include the area overhead of prior sparse microarchitectural enhancements: DS-STC~\cite{wang2021dual}, RM-STC~\cite{huang2023rm}, and Coruscant-STC~\cite{joo2025coruscant}. The \name design is consistently the most area-efficient, owing to its lightweight RLC decoder and the reuse of existing register files for accumulation rather than introducing dedicated compute structures.

\begin{table}[h]
\centering
\vspace{-2.5mm}
\caption{Area overhead comparison to prior works. Available
numbers are taken from respective paper. All area units are in mm$^2$. Overhead relative to full GPU is shown in parentheses for \name.}
\vspace{-2.5mm}
\begin{adjustbox}{max width=0.85\linewidth}
\begin{tabular}{l|c|c|c|c}
\toprule
\textbf{GPU} & \textbf{DS}~\cite{wang2021dual} & \textbf{RM}~\cite{huang2023rm} & \textbf{Coruscant}~\cite{joo2025coruscant}&\textbf{Celty-SIMT} \\
\midrule
V100 (815) & 12.85 & -- & 0.51  & 0.22 (\textbf{0.03\%})\\
A100 (826) & -- & 14.87  & 0.68 & 0.33 (\textbf{0.04\%})\\
H100 (814) & -- & -- & 1.44& 0.68 (\textbf{0.08\%})\\
\bottomrule
\end{tabular}
\vspace{-5mm}
\end{adjustbox}
\label{tab:area_power}
\end{table}

\begin{figure}[h]
\centering
\includegraphics[width=0.9\linewidth]{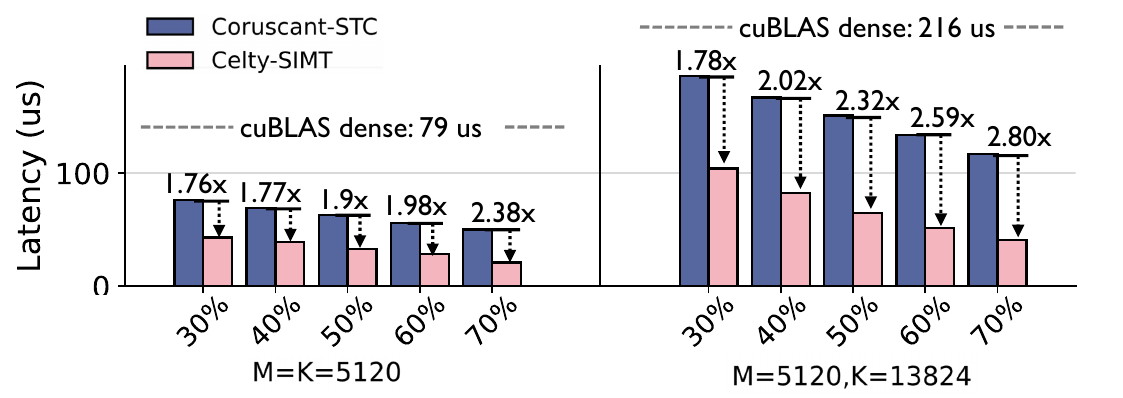}
\vspace{-3mm}
\caption{Comparison of sparse microarchitectural enhancements on 5120$\times$5120 and 5120$\times$13824 layers.}
\label{fig:hw_comparison}
\vspace{-4mm}
\end{figure}

\textbf{Performance comparison.} Figure~\ref{fig:hw_comparison} compares the \name Sparse SIMT Core against Coruscant's sparse tensor core~\cite{joo2025coruscant} on single-user decoding workloads. While Coruscant targets SpMM with $N\geq 8$ and decompresses sparse data into dense tiles before tensor-core execution, \name operates directly on compressed RLC-CSC data at N=1. Across all evaluated layers and sparsity levels, \name achieves on average 2.13$\times$ speedup over Coruscant-STC, with gains reaching up to 2.8$\times$ at 70\% dual-sparsity.

\begin{figure}[h]
\centering
\includegraphics[width=0.75\linewidth]{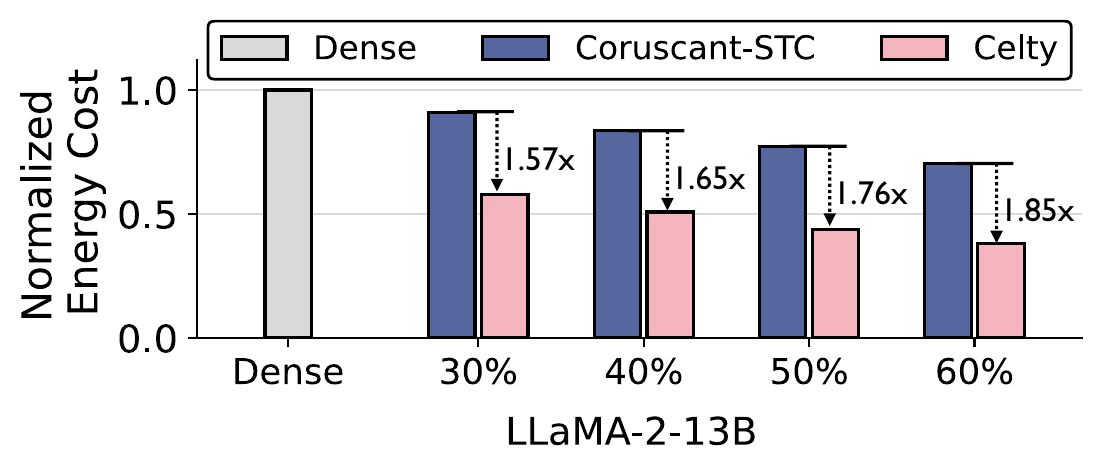}
\vspace{-2.5mm}
\caption{Energy cost of different GPU kernels (normalized to cuBLAS on A5000), each running on its own co-design-enhanced GPU.}
\label{fig:energy_comparison}
\vspace{-2.5mm}
\end{figure}

\textbf{Energy Efficiency comparison.} For energy efficiency, we report the normalized energy cost comparison between the dense cuBLAS kernel running on A5000, Coruscant-STC kernel running on Coruscant-STC enhanced A5000, and \name kernel running on \name-Sparse-SIMT enhanced A5000. As shown in Figure~\ref{fig:energy_comparison}, at 60\% dual-sparsity, \name reduces the energy cost by 1.85$\times$ compared to Coruscant-STC. Energy efficiency of \name comes from the kernel's speedup at minimal cost (on A5000, \name Sparse SIMT's total power overhead is 88.90\,mW, which is 0.04\% of the TDP). 

\subsection{Ablation Study}
\label{sec:ablation}

\begin{figure}[t]
\centering
\includegraphics[width=0.85\linewidth]{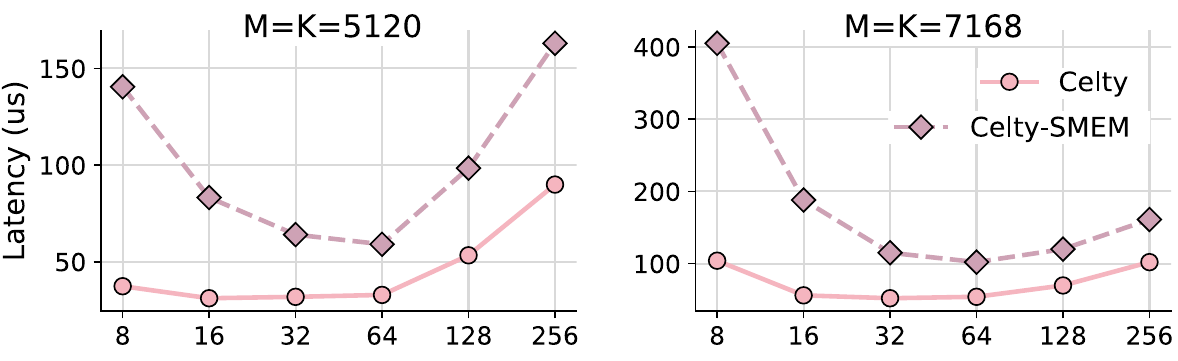}
\vspace{-2.5mm}
\caption{Kernel latency vs.\ Split-K parallelism at 50\% dual-sparsity on 5120$\times$5120 and 7168$\times$7168 layers.}
\vspace{-2mm}
\label{fig:ablation_parallelism}
\end{figure}

\textbf{Effect of Split-K parallelism.}
We study the impact of the inner-K-loop size on kernel performance by varying the Split-K parallelism. Figure~\ref{fig:ablation_parallelism} shows kernel latency across configurations for both 5120$\times$5120 and 7168$\times$7168 layers at 50\% dual-sparsity. Increasing Split-K parallelism (i.e., decreasing the inner-K-loop size) reduces latency up to an inner-loop size of 64, beyond which the cost of global atomic additions for partial-sum merging dominates. We fix the inner-K-loop size to 64 for all other experiments.

\begin{figure}[h]
\centering
\includegraphics[width=0.85\linewidth]{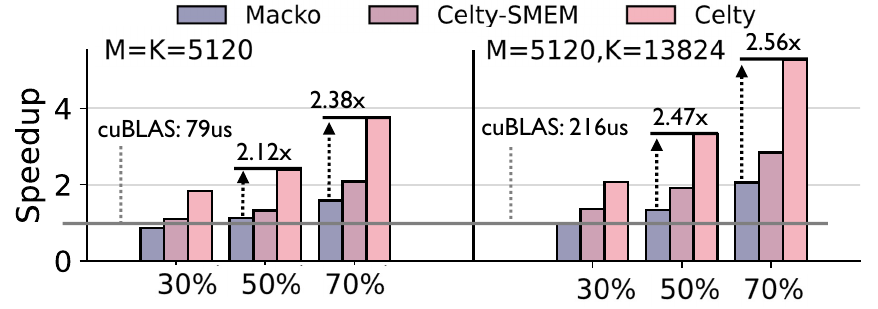}
\vspace{-2.5mm}
\caption{Comparison between \name and Macko across sparsity levels and layer dimensions.}
\vspace{-2.5mm}
\label{fig:ablation_macko}
\end{figure}

\textbf{Comparison with Macko across sparsity.}
Macko~\cite{macko2025macko} is an SpMV kernel that operates on an RLC-CSR format using SIMT cores. However, its execution flow dedicates each warp to a single weight row, an approach optimized for weight-only sparsity that cannot exploit activation sparsity. For a comprehensive evaluation, we provide the performance comparison of \name against Macko across sparsity levels and layer dimensions in Figure~\ref{fig:ablation_macko}. Both the SMEM-based \name kernel and the \name Sparse SIMT Core outperform Macko across all configurations, with the dual-sparsity advantage growing at higher sparsity: \name achieves on average 2.3$\times$ and 2.5$\times$ speedup over Macko at 50\% and 70\% dual-sparsity, respectively.

\textbf{Accuracy under dual-sparsity.}
Table~\ref{tab:ablation_accuracy_ppl} reports WikiText-2 perplexity for LLaMA-2-7B and LLaMA-2-13B under different sparsity methods at 50\% sparsity. We also report the average accuracy across three downstream tasks using LM Eval Harness~\cite{gao2021framework}: 5-shot MMLU~\cite{hendrycks2020measuring}, 0-shot PiQA~\cite{bisk2020piqa}, and 0-shot Winogrande~\cite{sakaguchi2021winogrande}. Structured pruning (ShortGPT) and semi-structured 2:4 pruning (SparseGPT) both incur severe perplexity degradation. Unstructured pruning (SparseGPT) preserves accuracy best, while dual-sparsity (DuoGPT) achieves 8.58 (7.17) perplexity on LLaMA-2-7B (13B), a moderate increase over the dense baseline of 5.47 (4.88) that confirms the sparsity levels targeted by \name remain within an acceptable accuracy budget.

\begin{table}[t]
  \centering
  \caption{Perplexity and downstream accuracy under different sparsity methods at 50\% model sparsity.}
  \vspace{-2.5mm}
  \begin{adjustbox}{max width=0.85\linewidth}
    \begin{tabular}{l||c|c|c}
    \toprule\toprule
    \textbf{Methods (Sparsity type)} & \textbf{LLM Model} &\textbf{Wiki2$(\downarrow)$} & \textbf{Avg Acc$(\uparrow)$} \\
    \midrule
    Dense &LLaMA-2-7B &5.47 & 63.3\\
    SparseGPT (50\% unstructured) &LLaMA-2-7B & 6.51& 60.1\\
    SparseGPT (2:4) &LLaMA-2-7B& 12.1 & 53.4\\
    ShortGPT (50\% structured) &LLaMA-2-7B & 368.8 & 47.2\\
    \rowcolor{sakura!70}DuoGPT (50\% dual-sparsity) &LLaMA-2-7B& 8.58 & 55.7\\
    \midrule
    Dense &LLaMA-2-13B & 4.88 &68.3\\
    SparseGPT (50\% unstructured) &LLaMA-2-13B & 5.63 & 64.8\\
    SparseGPT (2:4) &LLaMA-2-13B& 9.56 &  56.9 \\
    ShortGPT (50\% structured) &LLaMA-2-13B & 191.3 & 48.0\\
    \rowcolor{sakura!70}DuoGPT (50\% dual-sparsity) &LLaMA-2-13B& 7.17& 59.7\\
    \bottomrule\bottomrule
    \end{tabular}
  \end{adjustbox}
\label{tab:ablation_accuracy_ppl}
\end{table}

\begin{figure}[h]
\centering
\includegraphics[width=0.85\linewidth]{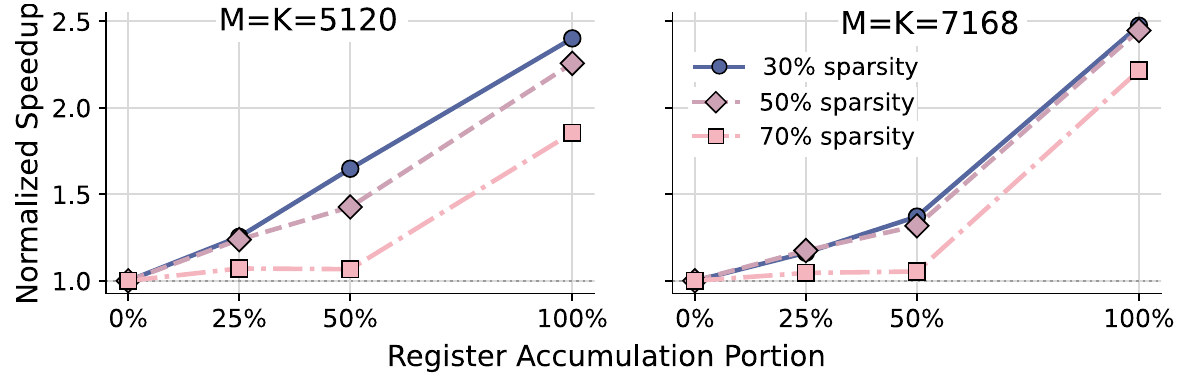}
\vspace{-2.5mm}
\caption{Normalized speedup of \name Sparse SIMT Core as the register-accumulation portion varies from 0\% (pure SMEM) to 100\% (pure register) across dual-sparsity levels.}
\vspace{-2.5mm}
\label{fig:ablation_reg}
\end{figure}

\textbf{Register vs.\ SMEM accumulation tradeoff.}
We vary the register-accumulation portion on \name sparse SIMT kernel from 0\% (pure SMEM) to 100\% (pure register) in Figure~\ref{fig:ablation_reg}. The gains are largest at low dual-sparsity, where denser scatter updates intensify SMEM bank conflicts: on a 5120$\times$5120 layer at 30\% dual-sparsity, only 50\% register accumulation can yield 1.65$\times$ speedup over pure SMEM accumulation, reaching 2.40$\times$ at 100\%. At 70\% dual-sparsity, conflicts are less frequent and 50\% register allocation provides only 1.07$\times$ improvement, yet full register accumulation still achieves 1.86$\times$, confirming that SMEM access latency remains a bottleneck beyond conflicts alone. Crucially, this study shows that fully register-resident accumulation is not required—the partition ratio can be tuned to balance speedup against occupancy constraints.

\section{Conclusion}

We presented \name, a co-designed sparse format, GPU kernel, and SIMT microarchitecture for dual-sparse LLM inference, formulated as SpMSpV operations. The \name Sparse SIMT Core eliminates software reconstruction overhead and shared-memory bottlenecks through a hardware RLC decoder and register-file accumulation, achieving up to 5.3$\times$ speedup over cuBLAS with negligible area overhead and up to 2.63$\times$ end-to-end decode speedup on LLaMA-2-13B with bounded accuracy degradation.

\bibliographystyle{ACM-Reference-Format}
\bibliography{celty}

\end{document}